\documentclass[aps,prb,twocolumn,nopacs,10pt,a4paper,nofootinbib]{revtex4}
\usepackage[utf8]{inputenc}
\usepackage{graphicx,amsmath}
\usepackage{amssymb}
\usepackage{amsthm}
\usepackage{slashed}
\usepackage{braket}
\usepackage{bbold}
\usepackage[caption=false,singlelinecheck=off]{subfig}
\usepackage{booktabs}
\usepackage{blindtext}
\usepackage{bbm}
\usepackage[ pdftex, plainpages = false, pdfpagelabels, 
                 pdfpagelayout = useoutlines,
                 bookmarks,
                 bookmarksopen = true,
                 bookmarksnumbered = true,
                 breaklinks = true,
                 linktocpage,
                 colorlinks = true,  
                 linkcolor = blue,
                 urlcolor  = blue,
                 citecolor = blue,
                 anchorcolor = green,
                 hyperindex = true,
                 hyperfigures
                 ]{hyperref} 
\usepackage{enumerate}
\usepackage{tikz}
\usetikzlibrary{patterns}

\begin{document}
\title{Geometric Josephson effects in chiral topological nanowires}
\author{Christian Sp\r{a}nsl\"{a}tt}
\email{christian.spanslatt@fysik.su.se}

\affiliation{Department of Physics, Stockholm University, SE-106 91 Stockholm, Sweden}
\date{\today}
\begin{abstract}
One of the salient signatures of Majorana zero modes and topological superconductivity is a $4\pi$-periodic Josephson effect due to the combination of fermion parity conservation and the presence of a topologically protected odd number of zero energy crossings in the Andreev spectrum.

In this paper, we study this effect in Josephson junctions composed of two semiconducting nanowires with Rashba spin-orbit coupling and induced superconductivity from the proximity effect. For certain orientations of the external magnetic field, such junctions possess a chiral symmetry and we show how this symmetry allows the Andreev spectrum and the protected crossings to be shifted by introducing a relative angle between the two wires. The junction then displays a geometrically induced anomalous Josephson effect, the flow of a supercurrent in the absence of external phase bias. Furthermore, we derive a proportionality relation between the local current density and the local curvature for a single curved wire. This result can be viewed as a one-dimensional analogue of the recently proposed geo-Josephson effect [Kvorning \textit{et al.}, \href{https://arxiv.org/abs/1709.00482}{arXiv:1709.00482}]. Our two proposed effects can in principle be used as signatures of topological superconductivity in one dimension.

\end{abstract}
\pacs{03.65.Vf, 73.43.-f, 72.15.Nj}
\maketitle

\section{Introduction}
\label{sec:Introduction}
Majorana zero modes (MZMs) are particle-hole symmetric zero energy bound states predicted to appear on edges and topological defects in a certain class of superconductors (SCs)~\cite{Kitaev2001,Alicea2012}. Besides characterizing this novel state of matter, topological superconductivity, MZMs are also expected to exhibit non-Abelian statistics, which in turn is believed to be a promising route towards fault-tolerant topological quantum computation~\cite{Nayak2008}.

Much of the activity in current research on MZMs is to propose and design hybrid superconducting structures and experimental setups where the modes are expected to appear through suitable tuning of various parameters. Among these setups, one that has attracted particular attention is a one-dimensional (1D) semiconducting nanowire with strong spin-orbit coupling lying in proximity with an $s$-wave SC~\cite{Oreg2010,Lutchyn2010}. When such a wire is immersed in an external magnetic field above a certain critical field strength, the system is expected to enter a topological superconducting regime with MZMs exponentially localized at the two wire endpoints. 

The zero-energy and particle-hole symmetry properties of MZMs are further predicted to result in a robust quantized tunneling conductance of $2e^2/h$ at zero voltage bias due to perfect Andreev reflection\cite{Flensberg2010,Fidkowski2012,Diez2012,Setiawan2015}
 and also a $4\pi$-periodic Josephson current due to the conservation of fermion parity\cite{Kitaev2001,Kwon2004,Fu2009}. Deeply connected to this $4\pi$-Josephson effect is the appearance of an odd number of zero energy crossings in the Andreev spectrum when the junction phase difference is tuned. While the existence of an odd number crossings is guaranteed from the underlying particle-hole symmetry\cite{Lutchyn2010}, their specific locations in the spectrum are non-universal and will generally depend on microscopic details. 
  
In this paper, we consider Josephson junctions of the aforementioned nanowires with an additional chiral symmetry and show that a single crossing can be viewed as a phase transition between two topologically distinct regimes. Hence, both its existence and location in the spectrum are universal and can be obtained from topological arguments. We further exploit this observation to analyze junctions in which the Andreev spectrum is strongly affected by the system geometry. Our key observation is that, in the presence of chiral symmetry, the directional nature of Rashba spin-orbit coupling (RSOC) is inherited by the induced $p$-wave pairing in the topological regime. Consequently, two proximity induced wires connected with an offset angle can therefore exhibit an effective phase difference, despite the lack of external phase bias. Hence, such a configuration realizes a so-called $\phi_0$-junction, which has recently been analyzed in the context of topological superconductivity\cite{Dolcini2015,Nesterov2016,Huang2017}.

We also show that by introducing local curvature $\kappa(w)$ into a single nanowire parametrized by the coordinate $w$, there exists an explicit relation between the superconducting current density and the curvature
\begin{equation}
	\label{eq:IntroEquation}
	\langle J(w) \rangle = \frac{\rho(w)}{2m_\text{eff}} \kappa (w),
\end{equation}
where $\rho(w)$ is the local charge density along the wire and $m_\text{eff}$ is the effective electron mass in the topological regime. Similar ideas, that curvature may induce currents (and accompanying magnetic fields) in chiral superconductors have recently been proposed in 2D\cite{Kvorning2017} and we argue that the effects presented in this paper are of similar origin.

The remainder of this paper is organized as follows. We introduce our model of a 1D topological SC in Sec.~\ref{sec:Model} and discuss its symmetries. In Sec.~\ref{sec:GeoJosephson} ,we consider this model in the setting of Josephson junctions and show how the junction geometry affects the Andreev spectrum. These results are confirmed numerically in Sec.~\ref{sec:Numerics}. In Sec.~\ref{sec:Curvature}, we use the previous analysis to derive a proportionality relation between the curvature and current density for a single curved wire. We end with a summary and a discussion in Sec.~\ref{sec:Summary}. Throughout the paper we use units where $e=\hbar=1$ and we also assume zero temperature.

\section{Model of a 1D topological superconductor}
\label{sec:Model}
\subsection{Hamiltonian}
\label{sec:Hamiltonian}
Our starting point is a single straight 1D semi-conducting nanowire lying in the $x$-$y$ plane. The wire has a RSOC with strength $\alpha_R$ and lies in proximity with an $s$-wave SC with order parameter $\Delta$. The wire is also immersed in an external magnetic field $\mathbf{B}$ which defines the Zeeman field $\mathbf{h}\equiv \frac{1}{2}g\mu_B \mathbf{B}$, with $g$ the effective g-factor in the wire and $\mu_B$ being the Bohr magneton. 

Assuming the wire to be thin, so that only a single channel is occupied, we model this system with a BdG Hamiltonian acting on basis spinors
$\psi(w) = (u^{}_\uparrow(w),u^{}_\downarrow(w),v^{}_\uparrow(w),v^{}_\downarrow(w))^T$, where $w$ is the coordinate along the wire, $u$ and $v$ represents electron and hole components and $\uparrow,\downarrow$ refers to the spin-projection along the $z$-axis. The Hamiltonian reads
\begin{subequations}
\label{eq:FullNanowire}
\begin{align}
\label{eq:NanoWire}
&\mathcal{H}_\text{BdG}(p_w) = \begin{pmatrix}
h(p_w) & h_\Delta \\
h^\dagger_\Delta & -h^T(-p_w)
\end{pmatrix},\\
& h(p_w) = \frac{p_w^2}{2m^*}-\mu -\alpha_R p_w (\sigma_y\cos\varphi -\sigma_x\sin\varphi) + \mathbf{h} \cdot \boldsymbol{\sigma}, \label{eq:normalHam} \\
& h_\Delta = |\Delta|e^{i\phi_s}(i\sigma_y).
\end{align}
\end{subequations}
In these expressions, $\mathbf{p}_w=p_w(\cos \varphi,\sin \varphi)$ is the planar momentum operator parametrized by $\varphi$, the angle between the wire and the positive $x$-axis. The effective electron mass is denoted by $m^*$ and $\mu$ is the chemical potential. The RSOC is assumed to arise from some electrical field pointing in the $\mathbf{\hat{z}}$ direction so that it favours spin-alignment along some vector lying in the $x$-$y$ plane. This direction will be referred to as the RSOC direction and can in principle depend on $w$. We decompose the proximity induced SC order parameter as $|\Delta|e^{i\phi_s}$ with the phase parameter inherited directly from the underlying $s$-wave SC. The set of Pauli-matrices $\boldsymbol{\sigma}=\{\sigma_x,\sigma_y,\sigma_z\}$ act in spin space, and for later convenience we also define similarly the set of particle-hole Pauli matrices $\boldsymbol{\tau}=\{\tau_x,\tau_y,\tau_z\}$. The accompanying $2\times2$ unit matrices are denoted $\sigma_0$ and $\tau_0$ respectively.

It has been shown\cite{Oreg2010,Lutchyn2010,Halperin2012} that the Hamiltonian~\eqref{eq:FullNanowire} can be mapped onto a spinless $p$-wave SC model with a topological regime hosting edge MZMs \cite{Kitaev2001}.
This topological regime occurs when two conditions on $\mathbf{h}$ are met\cite{Osca2014,Rex2014}: 

\begin{enumerate}[(i)]
	\item The full field must satisfy $|\mathbf{h}|>h_c \equiv \sqrt{|\Delta|^2+\mu^2}$. Furthermore, if the Hamiltonian~\eqref{eq:FullNanowire} is put on a lattice (with unit lattice constant) which limits the spectrum, there is an additional upper critical field $\tilde{h}_c\equiv\sqrt{|\Delta|^2+(\mu-4t)^2}$, where $t\equiv\frac{1}{2m^*}$ is the hopping parameter. In that case, the full field must also satisfy $|\mathbf{h}|<\tilde{h}_c$.
	\item The projection of the Zeeman field onto the direction of the RSOC vector, $\mathbf{h}_\alpha$, must obey $|\mathbf{h}_\alpha| < |\Delta|$, otherwise the energy gap closes, and the system becomes metallic.
\end{enumerate}

\subsection{Symmetries}
\label{sec:Symmetries}
The BdG Hamiltonian \eqref{eq:FullNanowire} belongs generally to symmetry class $\mathcal{D}$\cite{Schnyder2008,Ryu2010,Kitaev2009} with a single antiunitary symmetry, the particle-hole symmetry $\mathcal{P}\mathcal{H}_\text{BdG}(p_w)\mathcal{P}^{-1} = -\mathcal{H}_\text{BdG}(-p_w)$, $\mathcal{P}^2=+1$. In our basis $\mathcal{P}=\tau_x \mathcal{K}$, with $\mathcal{K}$ being the complex conjugation operator. 

However, with perpendicular or parallel orientations of the Zeeman field; $\mathbf{h}=h\mathbf{\hat{z}}$ or $\mathbf{h}=h \left(\cos\varphi\mathbf{\hat{x}}+ \sin \varphi\mathbf{\hat{y}}\right)$ respectively, the model has an additional unitary chiral symmetry $\mathcal{C}\mathcal{H}_\text{BdG}(p_w)\mathcal{C}^\dag = -\mathcal{H}_\text{BdG}(p_w)$ and the system belongs to symmetry class $\mathcal{BDI}$ characterized by a topological winding number invariant~\cite{Schnyder2008,Ryu2010,Kitaev2009,Diez2012,Tewari2012,VolovikBook}. In a translationally invariant (with compact Brillouin zone) and gapped system, this invariant can be computed as\cite{Tewari2012}
 \begin{equation}
 \label{eq:windingnumber}
 	\nu \equiv \frac{1}{\pi i} \int_0^\pi d\theta(p_w) \in \mathbbm{Z},
 \end{equation}
 which cannot change as long as the energy gap is maintained. Here, $\theta(p_w)$ is the phase of $\text{det}A(p_w)$ and $A(p_w)$ is the off-diagonal block of the BdG Hamiltonian in the chiral basis
 \begin{equation}
\mathcal{H}^\mathcal{C}_{\text{BdG}}(p_w) = \begin{pmatrix}
 		0 & A(p_w) \\
 		A^T(-p_w) & 0
 	\end{pmatrix}.
 \end{equation}
The particular model \eqref{eq:FullNanowire}, realizes only $\nu \in \{0,-1,+1\}$, where $\nu=0$ is the trivial regime and $\nu = \pm 1$ are topological regimes with edge MZMs. Crucially, in these regimes $\nu$ typically assigns wires with SC phases that are even or odd multiples of $\pi$ with opposite winding numbers. Consequently, the chiral wires has a notion of directionality in the pairing. This feature is particularly visible in Andreev reflection, where low energy electron-hole and hole-electron scattering processes are phase shifted by $\pi$ since they experience the SC gap with different signs due to their relative direction\cite{Tanaka1995,Beenakker2012} (see also Ref.~\onlinecite{Spanslatt2017} for a recent discussion).

\section{Geometrical Josephson junctions}
\label{sec:GeoJosephson}
It follows that if two chiral 1D topological SCs with a phase difference of $\pi$ (mod $2\pi$) are connected in a Josephson junction, the junction must host an interface where the gap closes, since $\nu$ necessarily changes\cite{Spanslatt2015}. On this interface, a number of zero energy bound states, related to the difference in $\nu$, appear. The Majorana nature of these states follows from particle-hole symmetry. Note also that in class $\mathcal{BDI}$, both the existence and the location of the crossing is topologically protected, and it will therefore not change in the presence of symmetry respecting perturbations and is also independent on whether the junction is in the short or long junction limit. 

We therefore consider the model~\eqref{eq:FullNanowire} in the context of Josephson junctions for two different configurations. First, the Zeeman field is taken perpendicular to the junction, and secondly in the same plane as the junction.
\subsection{Junctions with a perpendicular Zeeman field}
\label{sec:PerpJunctions}
\begin{figure}[t]
\captionsetup[subfigure]{position=top,justification=raggedright}
\subfloat[][]{
\includegraphics[width=0.96\columnwidth]{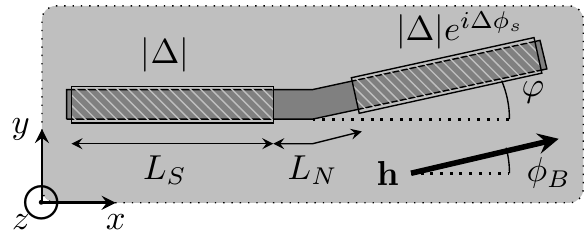}
\label{fig:SetupA}}
\\
\subfloat[][]{
\includegraphics[width=0.96\columnwidth]{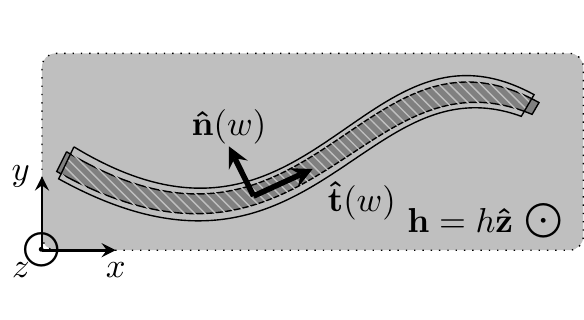}
\label{fig:SetupB}}
\label{fig:Setups}
\caption{(a) Schematics of a ``hinge'' type of Josephson junction with phase difference $\Delta \phi_s$, where the two wires are offset by the angle $\varphi$. The Zeeman field $\mathbf{h}$ is either long the $z$-direction or in the $x$-$y$ plane parametrized by the angle $\phi_B$. (b) Proximitized and curved wire in perpendicular Zeeman field. The wire is parametrized by a local Frenet frame $\mathbf{\hat{t}}(w)$-$\mathbf{\hat{n}}(w)$ defining the local curvature as $\partial_w\mathbf{\hat{n}}(w)=-\kappa(w)\mathbf{\hat{t}}(w)$. In the topological regime, the curvature induces an effective phase gradient $\kappa(w) = \partial_w\phi_p(w)$.}
\end{figure}

We consider a ``hinge'' type of junction where wire $1$ lies along the $x$-axis: $\mathbf{\hat{w}}_1=(1,0)$ of which the end point is coupled to wire $2$ lying along the direction $\mathbf{\hat{w}}_2=(\cos\varphi,\sin\varphi,0)$, see Fig.~\ref{fig:SetupA}. In addition, the wires lie in proximity to two separate $s$-wave SCs with externally controllable phases $\phi_{s1}$ and $\phi_{s2}$ respectively. The Zeeman field is taken as $\mathbf{h}=h\mathbf{\hat{z}}$, and we also choose a phase convention such that $\phi_{s1}=0$ and $\phi_{s2}\equiv\Delta \phi_s$: the $s$-wave phase difference between the wires. 

Considering first wire $1$ in isolation, the chiral symmetry operator is given by $\mathcal{C}=\tau_x$. For wire $2$ in isolation, the unitary transformation $U=\text{diag}(e^{i \varphi},1,e^{-i \varphi},1)$ transfers the angle $\varphi$ to become a contribution to $\Delta\phi_s$. To satisfy the same chiral symmetry condition as wire 1 (that is using $\mathcal{C}=\tau_x$ which is required for consistency if the wires are to be coupled), we find that the second wire must fulfill the condition $\Delta\phi_{s}+\varphi = n\pi$, where $n \in \mathbbm{Z}$. Therefore, for $n$ odd, we then see by using Eq.~\eqref{eq:windingnumber} that the two wires have opposite winding numbers at this instance. Hence, if the wires are connected, there must exist a gap closing, a crossing, in the Andreev spectrum precisely at 
\begin{equation}
\label{eq:crossingcondition}
\Delta \phi_s = \pi-\varphi \quad \text{mod}\;2\pi.
\end{equation}
It then follows that by varying $\varphi$, the whole spectrum can be shifted because of the topological protection and $2\pi$-periodicity of the crossing.
 
To verify this reasoning further we now review the mapping onto the effective low energy $p$-wave superconductor. To this end, we construct the following eigen-spinors of the normal state Hamiltonian \eqref{eq:normalHam} as
\begin{subequations}
\label{eq:electronspinors}
\begin{align}
&	\ket{u_+(p_w)} = e^{-i\pi/4}\begin{pmatrix}
		e^{-i\varphi}\cos\frac{\theta_p}{2}  \\
		-i\sin\frac{\theta_p}{2}
	\end{pmatrix}, \\
	&\ket{u_-(p_w)} =  e^{-i\pi/4}\begin{pmatrix}
		-e^{-i\varphi}\sin\frac{\theta_p}{2} \\
		-i\cos\frac{\theta_p}{2} 
	\end{pmatrix}, 
\end{align}
\end{subequations}
where $\tan\theta_{p}=|\alpha p_w/h|$. By particle-hole symmetry, the corresponding hole spinors read
\begin{subequations}
\label{eq:holespinors}
\begin{align}
	& \ket{v_+(p_w)} = e^{i\pi/4}\begin{pmatrix}
		-e^{i\varphi}\cos\frac{\theta_p}{2}  \\
		i\sin\frac{\theta_p}{2}
	\end{pmatrix}, \\
	&\ket{v_-(p_w)} = e^{i\pi/4}\begin{pmatrix}
		e^{i\varphi}\sin\frac{\theta_p}{2} \\
		i\cos\frac{\theta_p}{2} 
	\end{pmatrix}. 
\end{align}
\end{subequations}
The electron and hole energies are $\pm E_\pm(p_w) = \frac{p_w^2}{2m^*}-\mu\pm\sqrt{\alpha_R^2p_w^2+h^2}$ respectively. In the strong magnetic field regime, $h\gg \alpha_R$, the spins are almost completely polarized within each band, and if $\mu$ lies in the gap between the two electron bands, one can project the full Hamiltonian~\eqref{eq:NanoWire} onto the bands $\ket{u_-(p_w)}$ and $\ket{v_-(p_w)}$. The result is an effective spin-less $p$-wave Hamiltonian\cite{Halperin2012,Alicea2011}
 \begin{equation}
 \label{eq:effectivepwave}
 	\mathcal{H}_p = \begin{pmatrix}
 		\frac{p_w^2}{2m_\text{eff}}-\mu_\text{eff} & p_w |\Delta_p| e^{i \phi_p} \\
 		p_w |\Delta_p| e^{-i \phi_p} & -\frac{p_w^2}{2m_\text{eff}}+\mu_\text{eff}
 	\end{pmatrix},
 \end{equation}
 where the effective parameters are $\mu_\text{eff}=\mu+h$, $\frac{1}{m_\text{eff}}=\frac{1}{m^*}(1-m^*\alpha_R^2/h)$, $|\Delta_p|= |\alpha_R \Delta/h|$. Most interestingly, in this basis, the effective $p$-wave phase is given by
\begin{equation}
\label{eq:phases}
\phi_p =  \phi_s + \varphi.
\end{equation}
For a single uniform wire, the extra geometrical phase contribution $\varphi$ is not important since it can be removed adjusting the overall phase factors on the spinors in \eqref{eq:electronspinors} and \eqref{eq:holespinors}. However, in the Josephson junction setup outlined above, the effective phase difference reads
\begin{equation}
	\label{eq:phasediff}
	\Delta \phi_p \equiv \phi_{p2}-\phi_{p1}= \Delta \phi_s + \varphi.
\end{equation}
In the effective model limit, there will be gap closing in the Andreev spectrum whenever $\Delta \phi_p=\pi$ (mod $2\pi$)\cite{Kwon2004,Spanslatt2015}, which by Eq.~\eqref{eq:phasediff} again leads to the condition~\eqref{eq:crossingcondition}.

In the short junction limit, where there are only two Andreev bound states, the spectrum and the zero temperature DC-current for a 1D $p$-wave Josephson junction were calculated in Ref.~\onlinecite{Kwon2004} (assuming fermion parity conservation). They read
\begin{subequations}
\label{eq:ABS}
\begin{align}
	 \epsilon_{\pm} &= \pm |\Delta_p|\sqrt{D}\cos(\frac{\Delta\phi_p}{2}), \\
	 I &=\sqrt{D}|\Delta_p| \sin(\frac{\Delta\phi_p}{2}),
	\end{align}
\end{subequations}
where $D$ is the junction transparency for a single channel.

By simply inserting the phase relation in Eq.~\eqref{eq:phasediff} into Eq.~\eqref{eq:ABS} we obtain a geometrically dependent spectrum and current
\begin{subequations}
\label{eq:ABSGeo}
\begin{align}
\label{eq:ABSGeo1}
	 \epsilon_{\varphi,\pm} &= \pm |\Delta_p|\sqrt{D}\cos(\frac{\Delta\phi_s + \varphi}{2}), \\ 
	 I_\varphi &= \sqrt{D}|\Delta_p|\sin(\frac{\Delta\phi_s + \varphi}{2}).
	\end{align}
\end{subequations}

In particular, there is a finite Josephson current, even if the externally controlled $\Delta \phi_s=0$. We denote this anomalous current as geometrical and define
\begin{equation}
\label{eq:GeoCurrent}
I_{Geo} \equiv I_{\varphi = 0}=\sqrt{D}|\Delta_p|\sin(\frac{\varphi}{2})	,
\end{equation}
which is one of the main results of this paper. $I_{Geo}$ is maximal for anti-parallel wires $\varphi = \pi$, which is equivalent to the wires having opposite signs of the Rashba coupling $\alpha_R$. The junction then realizes a topological $\pi$-junction~\cite{Ojanen2013,Klinovaja2015}. In essence, our proposal of the geometrical Josephson junction is a generalization of such junctions which are based on chiral symmetry.

\subsection{Junctions with a planar Zeeman field}
\label{sec:PlanarJunctions}
Next, we discuss the same setup as in the previous section, but with a planar Zeeman field $\mathbf{h}=h \left(\cos\phi_B\mathbf{\hat{x}}+ \sin \phi_B\mathbf{\hat{y}}\right)$, parametrized by the angle $\phi_B$.

As follows from the discussion in Sec.~\ref{sec:Symmetries}, the chiral symmetry will be broken for all values of $\Delta \phi_s$ unless the Zeeman field lies in parallel with both wires. If there is such parallel alignment, the crossing necessarily occurs at $\Delta \phi_s = \pi$ (mod $2\pi$)\cite{Lutchyn2010} and both the crossing and its location is topologically protected.

Accordingly, a finite $\varphi$ for wire $2$ will also break the chiral symmetry for all $\Delta \phi_s$, and we can not use winding number arguments for determining the location of the crossing in the Andreev spectrum. Still, the presence of a single crossing in the Andreev spectrum is protected by particle-hole symmetry\cite{Lutchyn2010}, but its location is non-universal. 

Indeed, for both wires along the $x$-axis ($\varphi=0$) and with a small $h_y=h\sin \phi_B$ , it was reported that location of the crossing occurs when $\Delta \phi_s = \pi+2\arcsin(h_y/|\Delta|)$ in the short junction limit~\cite{Dolcini2015,Nesterov2016,Huang2017}, due to a Fermi momentum mismatch mechanism. However, for longer junctions, the crossing location changes, indicating the non-universality (see Sec.~\ref{sec:Numerics}).

When $\mathbf{h}=h\mathbf{\hat{x}}$ and $\varphi$ is finite, we have found numerically (see Sec.~\ref{sec:Numerics}) that the crossing occurs when $\Delta\phi_s = \pi-\arcsin(h\sin\varphi/|\Delta|)$ in the short junction limit, but again, for longer junctions this is not the case. Nevertheless, the geometry of the junction affects the Andreev spectrum, but from a similar mismatch mechanism, rather than by changes in the condition for a topological phase transition as in Sec.~\ref{sec:PerpJunctions}. We leave the investigation of the combination of both an arbitrary planar Zeeman field and a finite ``hinge'' angle $\varphi$ to the future.

We do note however that due to point (ii) in Sec.~\ref{sec:Hamiltonian}, there is an upper bound on the Zeeman field component in the RSOC direction which restricts the range of possible $\varphi$ for the junction to be topological. This restriction is determined by the following expressions
\begin{subequations}
\label{eq:anglereq}
\begin{align}
\label{eq:CritangleA}
 \sin(\phi_{B,c}) &= \pm \frac{|\Delta|}{h}, \\
 \label{eq:CritangleB}
	 \sin(\phi_B-\varphi_c) &= \pm \frac{|\Delta|}{h}.
	\end{align}
\end{subequations}
In Eq.~\eqref{eq:CritangleA} $\phi_{B,c}$ is the maximally allowed angle for condition (ii) to be fulfilled for wire $1$. Given that $\phi_B<\phi_{B,c}$, Eq.~\eqref{eq:CritangleB} then limits the field projection on the $\varphi$-dependent RSOC vector in wire $2$ and thereby determines $\varphi_c$, the maximally allowed offset angle.

\section{Numerical results}
\label{sec:Numerics}
\begin{figure}[t]
\captionsetup[subfigure]{position=top,justification=raggedright}
\subfloat[Subfigure 1 list of figures text][]{
\includegraphics[width=0.48\columnwidth]{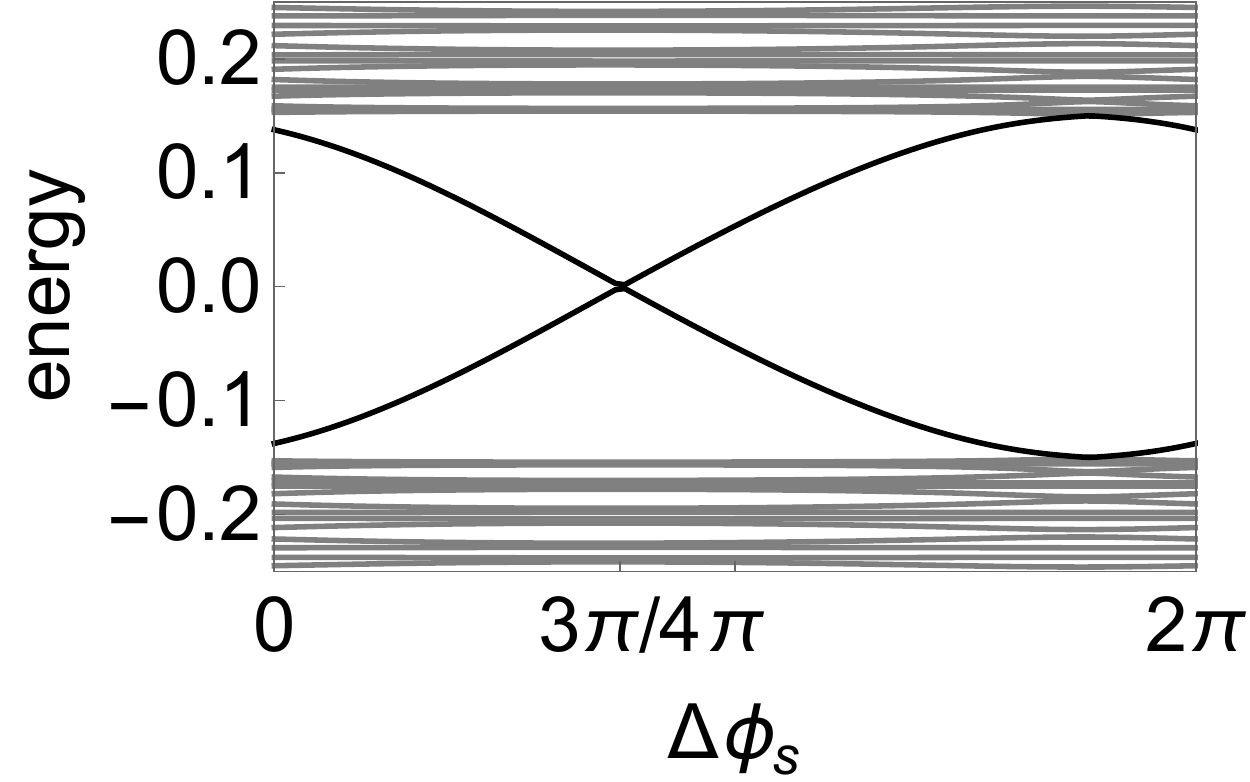}
\label{fig:Spectrum1}}
\subfloat[Subfigure 2 list of figures text][]{
\includegraphics[width=0.48\columnwidth]{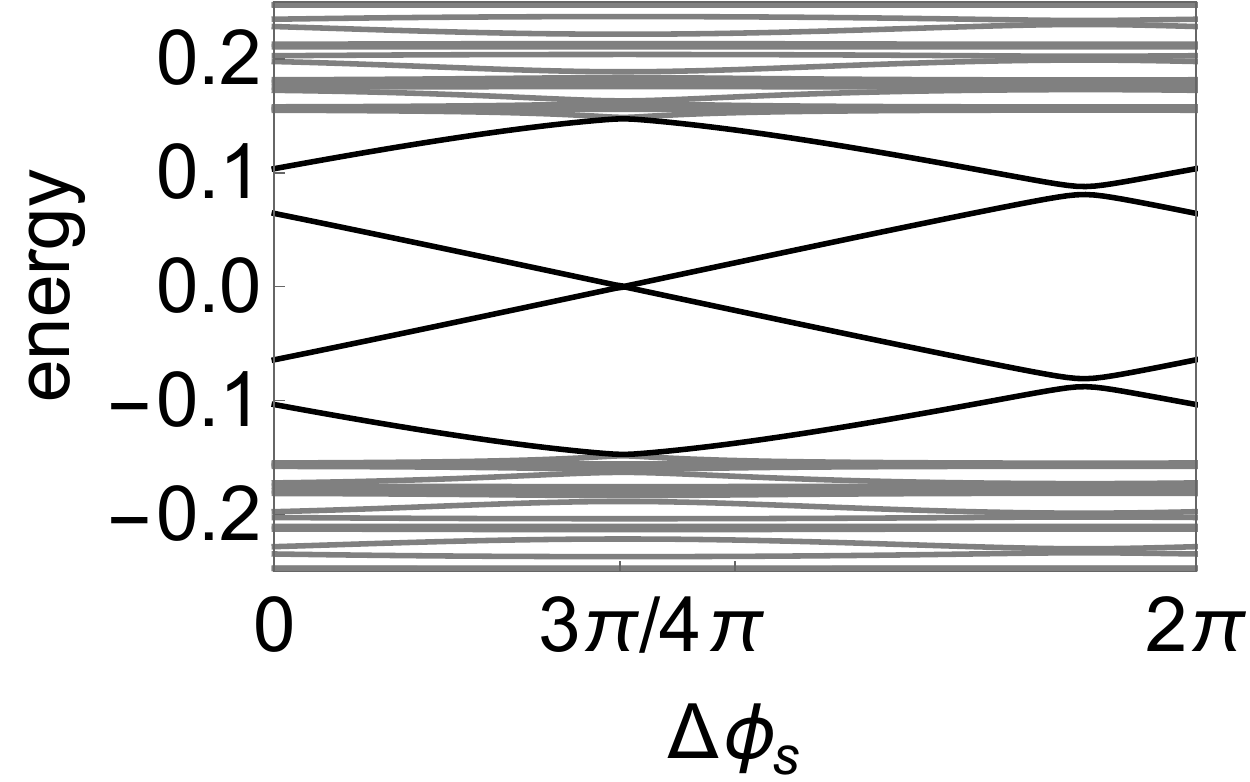}
\label{fig:Spectrum2}}
\label{fig:Spectrum12}
\caption{(a) Andreev (black) and bulk (gray) spectrum of a short, $L_N=0$, geometrical Josephson junction as a function of the phase difference $\Delta \phi_s$. Flat midgap states, corresponding to outer edge Majorana zero modes states have been removed for clarity. The parameters are $L_S=120$, $t=1.0$, $|\Delta|=1.0$, $\mu=0.0$, $h=3.0$ and $\alpha_R=0.5$. (b) Same as (a) but with $L_N=10$.}
\end{figure}

\begin{figure}[t]
\captionsetup[subfigure]{position=top,justification=raggedright}
\subfloat[Subfigure 1 list of figures text][]{
\includegraphics[width=\columnwidth]{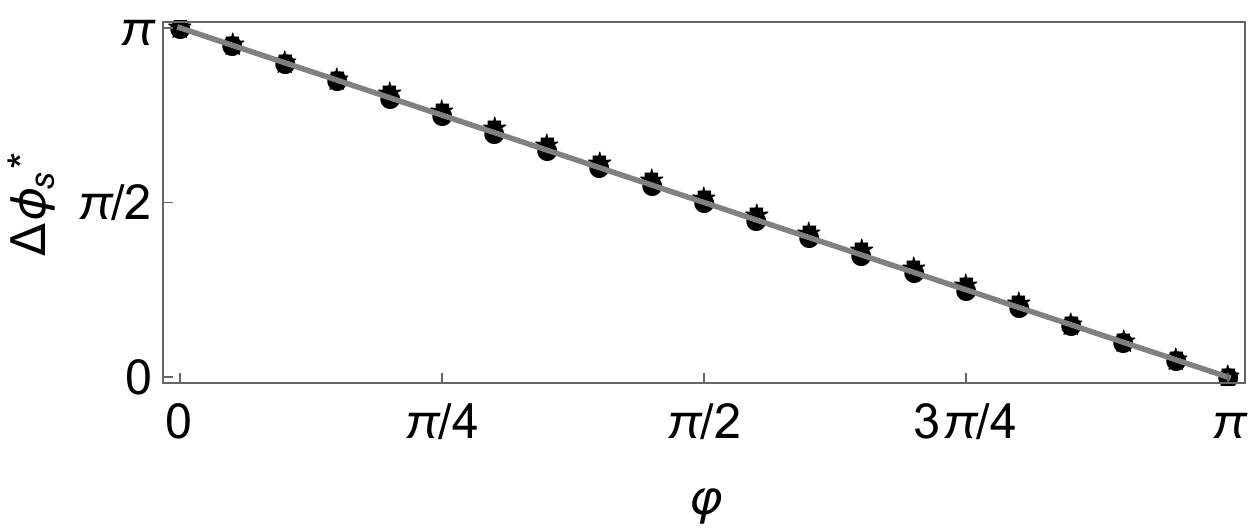}
\label{fig:SpectrumA}}
\\
\subfloat[Subfigure 2 list of figures text][]{
\includegraphics[width=\columnwidth]{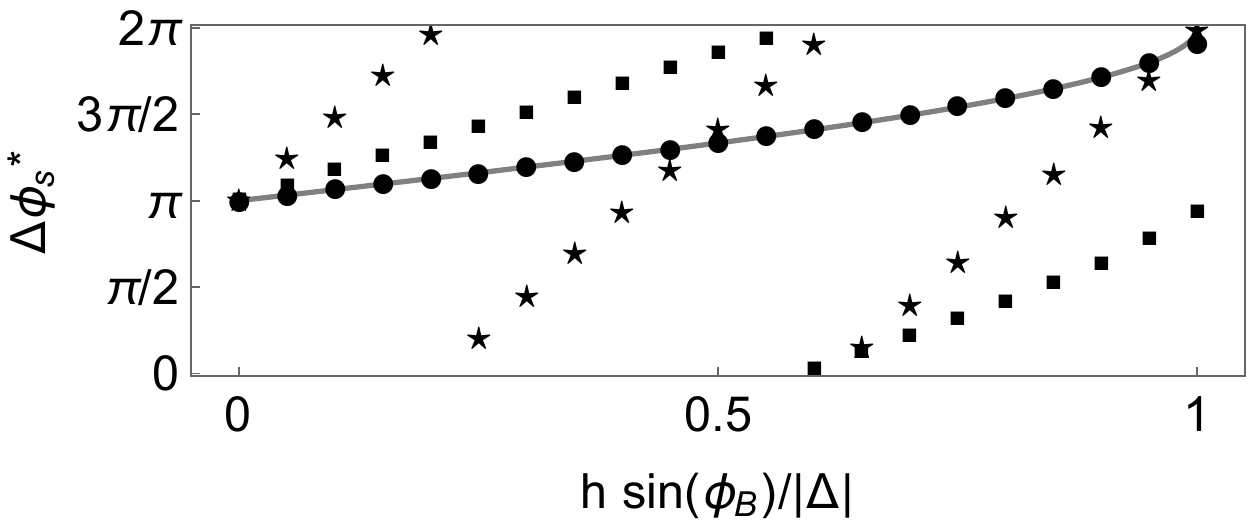}
\label{fig:SpectrumB}}
\\
\subfloat[Subfigure 2 list of figures text][]{
\includegraphics[width=\columnwidth]{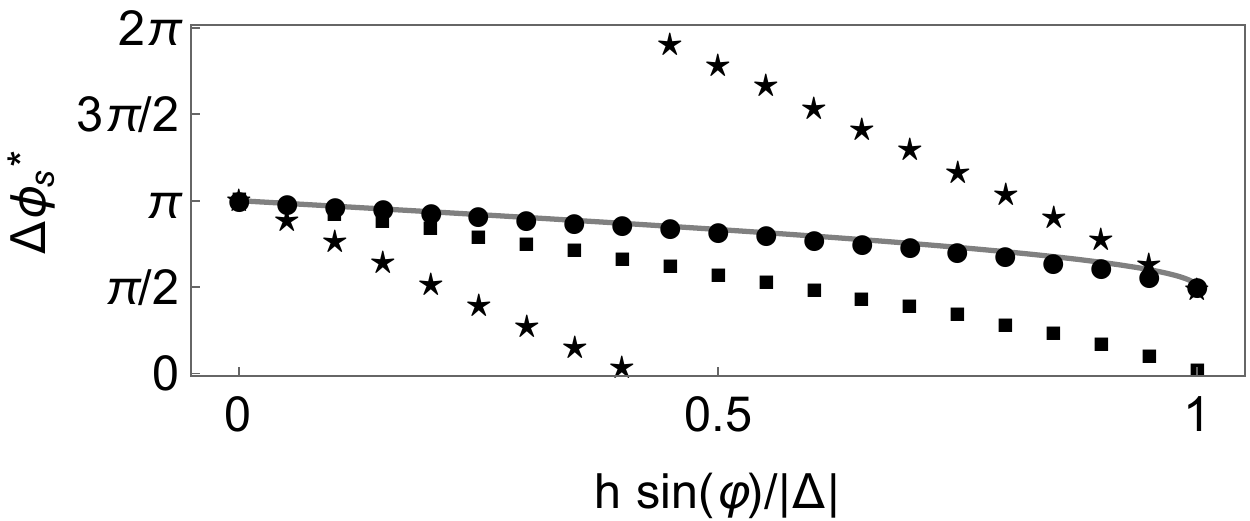}
\label{fig:SpectrumC}}
\label{fig:SpectrumABC}
\caption{Numerically calculated locations of zero energy crossings, $\Delta \phi_s^*$, in the $s$-wave phase difference $\Delta \phi_s$ for various configurations of the topological Josephson junction. In (a) $\mathbf{h}=h\mathbf{\hat{z}}$ and the gray solid line denotes $\Delta \phi_s^*=\pi-\varphi$, (b) $\mathbf{h}=h \left(\cos\phi_B\mathbf{\hat{x}}+ \sin \phi_B\mathbf{\hat{y}}\right)$, $\varphi=0$, and the gray line is $\Delta \phi_s^*=\pi+2\arcsin(h\sin(\phi_B)/|\Delta|)$, (c) $\mathbf{h}=h\mathbf{\hat{x}}$ and the gray line is $\Delta \phi_s^*=\pi-\arcsin(h\sin(\varphi)/|\Delta|)$. For all plots: $L_S=120$, $t=1.0$, $|\Delta|=1.0$, $\mu=0.0$, $h=3.0$ and $\alpha_R=0.5$. The normal segment lengths are $L_N=0$ (circles), $L_N=10$ (squares) and $L_N=40$ (stars).}
\end{figure}
To verify our analytical calculations, we next numerically compute the low energy spectrum of the various Josephson junction setups. To this end, we use a discretized version of~\eqref{eq:FullNanowire} (on a lattice with unit lattice constant) given by
\begin{align}
\label{eq:Tightbinding}
\mathcal{H} &= -t\sum_{j,\sigma} c^{\dag}_{j\sigma} c^{}_{j+1\sigma}+h.c.-(\mu-2t)\sum_{j,\sigma} c^{\dag}_{j\sigma} c^{}_{j\sigma}\notag\\
& + \sum_{j,\sigma, \sigma'} c^{\dag}_{j\sigma} \left(\mathbf{h} \cdot \boldsymbol{\sigma}\right)_{\sigma \sigma'} c^{}_{j\sigma'}\notag\\
&-\frac{i\alpha_R}{2}\sum_{j,\sigma, \sigma'} c^{\dag}_{j\sigma}\left(\sigma_y\cos\varphi_{j}-\sigma_x\sin\varphi_{j}\right)_{\sigma \sigma'} c^{}_{j+1\sigma'}-h.c. \notag \\
&+\sum_{j} \Delta_j c^{}_{j\uparrow}c^{}_{j\downarrow}+h.c.,
\end{align}
where $t\equiv 1/2m^*$ is the hopping parameter and $\varphi_{j}$ is the angle between site $j$ and the positive $x$ axis. We model the SC-normal-metal-SC Josephson junction setup by taking
\begin{equation}
\Delta_j =\begin{cases}
    |\Delta| &\text{for } 1\leq j\leq L_S \\
    ~0 &\text{for }  L_S < j \leq L_S+L_N \\
    |\Delta|e^{i \Delta \phi_s} &\text{for } L_S+L_N<j\leq 2L_S+L_N,
  \end{cases}
\end{equation}
where $L_S$ denotes the number of sites in each of the SC regions, and $L_N$ that of the normal region, see Fig.~\ref{fig:SetupA}. Throughout this section, we choose $L_S=120$, $t=1.0$, $|\Delta|=1.0$, $\mu=0.0$, $h\equiv|\mathbf{h}|=3.0$ and $\alpha_R=0.5$.

We first model the setup described in Sec.~\ref{sec:PerpJunctions} and accordingly we take $\mathbf{h}=h\mathbf{\hat{z}}$, wire 1 along the direction $\mathbf{\hat{w}}_1=(1,0)$ and wire 2 lies along $\mathbf{\hat{w}}_2=(\cos\varphi,\sin\varphi)$, see Fig.~\ref{fig:SetupA}. For concreteness, we also choose $\varphi=\pi/4$. The resulting spectra for $L_N=0$ and $L_N=10$ are depicted in Fig.~\ref{fig:Spectrum1} and~\ref{fig:Spectrum2} respectively. We note that the crossing in both plots occurs at $\Delta \phi_s = 3\pi/4$ in agreement with Eq.~\eqref{eq:crossingcondition}. Furthermore, the Andreev spectrum in Fig.~\ref{fig:Spectrum1} matches Eq.~\eqref{eq:ABSGeo1} very well with $D\approx 0.85$.

We next generalize this calculation to general angles $\varphi$ and for a few different junction lengths $L_N$. The result is shown in Fig.~\ref{fig:SpectrumA}. The location of the crossings, $\Delta \phi_s^*$, agree very well with Eq.~\eqref{eq:crossingcondition}, represented by the gray solid line, and does not change when increasing $L_N$. We attribute this feature to the topological nature of the crossing location. 

We repeat the previous calculation for the planar setup of Sec.~\ref{sec:PlanarJunctions}. We first take $\mathbf{h}=h \left(\cos\phi_B\mathbf{\hat{x}}+ \sin \phi_B\mathbf{\hat{y}}\right)$ and $\varphi=0$. The results are shown in Fig.~\ref{fig:SpectrumB} and are in agreement (gray solid line) with Refs.~\onlinecite{Dolcini2015,Nesterov2016,Huang2017} in the short junction limit $L_N=0$. For longer junctions, the crossing locations deviate significantly from this line, and indicates that the crossing locations are not protected by topology. We also note that in this limit, the crossings changes from below to above $\pi$ and vice versa, indicated by the discontinuities in Fig.~\ref{fig:SpectrumB} and Fig.~\ref{fig:SpectrumC}.

Finally, we choose $\mathbf{h}=h\mathbf{\hat{x}}$ and vary $\varphi$. The results are presented Fig.~\ref{fig:SpectrumC}. In the short junction limit, we find the approximate relation $\Delta\phi_s^* = \pi-\arcsin(h\sin\varphi/|\Delta|)$ which is violated for longer junctions. In all figures, we associate the slight deviations from the dashed lines to finite size effects. We have also checked that weak scalar disorder does not qualitatively change our results.


%
We conclude that our numerical calculations support the concept of geometrical Josephson junctions in Sec.~\ref{sec:GeoJosephson}.
\section{From curvature to phase gradient to current}
\label{sec:Curvature}
Given the result of mapping from Eq.~\eqref{eq:FullNanowire} onto the low energy theory~\eqref{eq:effectivepwave}, we anticipate that the induced phase shift in Eq.~\eqref{eq:phases} should influence a wire where $\varphi$ varies continuously, which is the situation for a curved wire, see Fig.~\ref{fig:SetupB}. We therefore move on to generalise the mapping onto the effective theory to a smoothly curved wire in the $x$-$y$ plane. The symmetrized and Hermitian normal state Hamiltonian in real space reads\cite{Ortix2015a,Ortix2015b,Ying2017}
\begin{equation}
	\label{eq:curvedRashba}
	h_\text{curved}(w) = -\frac{\partial^2_w}{2m^*}-\mu -\frac{\alpha_R}{2}\lbrace-i\partial_w , \sigma(w) \rbrace +  h \sigma_z,
\end{equation}
where $\lbrace\cdot,\cdot\rbrace$ denote the anti-commutator and $\sigma(w)$ is a local Pauli matrix co-moving along with the wire. Explicitly, $\sigma(w)=\boldsymbol{\sigma}\cdot \mathbf{\hat{n}}(w)$, where $\mathbf{\hat{n}}(w)$ is the local unit vector normal to the wire direction. The Frenet-Serret equations then define the local curvature, $\kappa(w)$, of the wire through $\partial_w\mathbf{\hat{n}}(w)=-\kappa(w)\mathbf{\hat{t}}(w)$ with $\mathbf{\hat{t}}(w)$ being the local unit tangent vector along the wire. Consequently, the RSOC favours spin-alignment in a direction determined by the local curvature of the wire according to
\begin{equation}
	\label{eq:curvature}
	\varphi(w) = \int^w_0\;dw' \kappa(w').
\end{equation}
By again following the steps of the mapping onto the topological superconductor, we obtain
\begin{equation}
 \label{eq:curvedpwave}
 	\mathcal{H}_p = \begin{pmatrix}
 		-\frac{\partial_w^2}{2m_\text{eff}}-\mu_\text{eff} & \frac{1}{2} \lbrace-i\partial_w , |\Delta_p| e^{i \phi_p(w)} \rbrace\\
 		\frac{1}{2}\lbrace-i\partial_w , |\Delta_p| e^{-i \phi_p(w)} \rbrace & \frac{\partial_w^2}{2m_\text{eff}}+\mu_\text{eff}
 	\end{pmatrix},
\end{equation}
and the local curvature in Eq.~\eqref{eq:curvature} enters the local $p$-wave pairing phase through
\begin{equation}
	\label{eq:curvaturephase}
	\phi_p(w) = \phi_s +\int^w_0\;dw' \kappa(w').
\end{equation}
Hence, the presence of curvature in the wire is equivalent to a phase gradient of the effective order parameter.

By using the unitary transformation $U=\exp(-i \phi_p(w)\tau_z/2)$, the phase $\phi_p(w)$ can be removed from the pairing term, and the transformed Hamiltonian $\mathcal{H}_p \mapsto U \mathcal{H}_p U^\dag$ becomes
\begin{equation}
 \label{eq:curvedpwavetransformed}
 	\mathcal{H}_p = \begin{pmatrix}
 		\frac{\left(-i\partial_w + \partial_w \phi_p/2\right)^2}{2m_\text{eff}}-\mu_\text{eff} & -|\Delta_p|i\partial_w ,  \\
		-|\Delta_p|i\partial_w & -\frac{\left(-i\partial_w - \partial_w \phi_p/2\right)^2}{2m_\text{eff}}+\mu_\text{eff}
 	\end{pmatrix},
\end{equation}
where it has been assumed that $|\Delta_p|$ is spatially constant. This expression makes manifest that the local curvature $\kappa(w) = \partial_w\phi_p(w)$ enters the Hamiltonian just as an electromagnetic vector potential, similar to the 2D geo-Josephson effect\cite{Kvorning2017}.

The charge current density operator $J(w)$ can be derived by first employing the usual minimal substitution $-i\partial_w \rightarrow -i\partial_w \pm A_w$ where the signs are different for the electron and hole components. Note however, that to ensure invariance under electromagnetic gauge transformations
\begin{subequations}
\begin{align}
	\label{eq:Gaugetransformation}
	A_w &\rightarrow A_w' = A_w+\partial_w \chi(w) \\
	\phi_p(w) &\rightarrow \phi_p'(w) = \phi_p(w)-2\chi(w),
\end{align}
\end{subequations}
the minimal substitution rule does not apply to the pairing terms. We then obtain the current density operator from $\langle J(w) \rangle \equiv \frac{\delta H_p^A}{\delta A(w)}$, where $H_p^A=\int dw \Psi^\dag(w) \mathcal{H}_p(w) \Psi(w)$ is the full $p$-wave Hamiltonian, as
\begin{equation}
	\label{eq:currentoperator}
	J(w) = \frac{1}{2m_\text{eff}}\left(-i \overset{\rightarrow}{\partial}_w + i\overset{\leftarrow}{\partial}_w \right)\tau_0 + \frac{\left(\partial_w \phi_p(w) -2A(w) \right)}{2 m_\text{eff}}\tau_z. 
	\end{equation}
 Assuming further that the the system is in its zero temperature ground state with no single particle excitations, we omit the first term in \eqref{eq:currentoperator} and obtain the supercurrent contribution
 \begin{equation}
	\label{eq:current3}
\langle J(w)\rangle = \frac{\rho(w)}{2m_\text{eff}} (\partial_w \phi_p(w)-2A(w)),
\end{equation}
where $\rho(w) = |u(w)|^2-|v(w)|^2$ is the local charge density along the wire in the ground state. Finally, setting $A(w)=0$ and using Eq.~\eqref{eq:curvaturephase}, the geometrically induced supercurrent density reads
\begin{equation}
	\label{eq:current2}
\langle J(w)\rangle = \frac{\rho(w)}{2m_\text{eff}} \kappa(w),
\end{equation}
which is Eq.~\eqref{eq:IntroEquation} in Sec.~\ref{sec:Introduction}. This effect is similar to that described in Ref.~\onlinecite{Klinovaja2015}, where the Rashba electrical field direction is changed constinously along a straight wire and thereby generates a phase gradient and a supercurrent in the topological regime. Here instead, the Rashba electrical field is fixed, but the wire direction changes continously, making the connection between current and curvature manifest and also relates the effect to a higher dimensional analogue~\cite{Kvorning2017}.

\section{Summary and Discussion}
\label{sec:Summary}
In this paper we considered Josephson junctions composed of 1D topological superconducting wires. We showed that it is possible to generate a contribution to the Josephson current which originates from a geometric offset angle between two wires. In particular, this contribution results in a geometrically induced anomalous Josephson current which can be traced to the directional nature of Rashba spin-orbit coupling inherited by the effective $p$-wave pairing in the topological regime. We also showed that this directional dependence manifests itself by inducing a phase gradient in the presence of curvature. Since a superconductor responds to a phase gradient by generating  a supercurrent, we could derive an explicit current-curvature relationship for chiral 1D $p$-wave SCs.
 
Regarding the experimental feasibility, observing a curvature induced supercurrent through the curving of a planar nanowire is perhaps not within current technological range. In addition, setups with proximity coated nanowires and a perpendicular magnetic field configuration has the drawback of low critical magnetic fields, which complicates the transition into the topological regime~\cite{Higginbotham2015}.  A more reasonable proposal would instead be to use etched narrow channels in a two-dimensional electron gas with strong spin-orbit coupling\cite{Hell2017}. Such a system, immersed in a thin semiconducting layer sandwiched between a conventional $s$-wave superconductor and a ferromagnetic insulator could realize the model \eqref{eq:FullNanowire} with $\mathbf{h}=h\mathbf{\hat{z}}$~\cite{Sato2010,Black-Schaffer2011}. 
 
Going beyond the single channel description of the wires, it has been shown\cite{Tewari2012,Tewari2012b,Diez2012} that the chiral symmetry is still approximately present as long as the wire width is smaller than the spin-orbit length $l_{so}=(m^*\alpha_R)^{-1}$.

As a final remark, we note that the geometrical supercurrent \eqref{eq:IntroEquation} should be thought of as \textit{extrinsically} induced in the sense that it originates from curvature in the embedding 2D space, since no intrinsic geometry can be defined on a 1D line. That the 1D confined electrons still experience this extrinsic curvature is because of the Rashba effect, through which the electron spin introduces the notion of direction. This feature stands in contrast to the recently proposed geo-Meissner and geo-Josephson effects for 2D chiral $p$-wave superconductors\cite{Kvorning2017}, where the geometrical contribution to the Josephson current stems from the \textit{intrinsic} 2D curvature. Nonetheless, both effects arise because of a directional or chiral nature of the $p$-wave pairing through the electron spin degree of freedom.

\section*{ACKNOWLEDGMENTS}
The author is very grateful to Eddy Ardonne, Thors Hans Hansson, Iman Mahyaeh and Axel Gagge for numerous helpful discussions and input. Thomas Kvorning, Jens Bardarsson, Ady Stern and Jorge Cayao are also acknowledged for useful comments.


%

\end{document}